\documentclass[a4paper,11pt]{article}
\usepackage{pos}

\title{Indirect detection, direct detection, and collider detection cross-sections for a 70 GeV dark matter WIMP}
\ShortTitle{cross-sections for a 70 GeV dark matter WIMP}

\author{Bailey Tallman}
\author{Alexandra Boone}
\author{Caden LaFontaine}
\author{Trevor Croteau}
\author{Quinn Ballard}
\author{Sabrina Hernandez}
\author{Spencer Ellis}
\author{Adhithya Vijayakumar}
\author{Fiona Lopez}
\author{Samuel Apata}
\author{Jehu Martinez }
\author*{Roland Allen}

\affiliation{Physics and Astronomy Department, Texas A\&M University}

\emailAdd{allen@tamu.edu}

\abstract{Assuming a dark matter fraction $\Omega_{DM} = 0.27$ and a reduced Hubble constant $h = 0.73$, we obtain a value of 70 GeV/c$^2$ for the mass of the dark matter WIMP we have previously proposed. We also obtain a value for the annihilation cross section given by $\langle \sigma_{ann} v \rangle = 1.19 \times 10^{-26} $ cm$^3$/s in the present universe, consistent with the current limits for dwarf spheroidal galaxies. Both the mass and cross-section are consistent with analyses of the Galactic-center gamma rays observed by Fermi-LAT and the antiprotons observed by AMS-02 if these data are interpreted as resulting  from dark matter annihilation. The spin-independent cross-section for direct detection in Xe-based experiments is estimated to be slightly above $10^{-48}$ cm$^2$, presumably just within reach of the LZ and XENONnT experiments with $\gtrsim 1000$ days of data taking. The cross-section for production in high-energy proton collisions via vector boson fusion is estimated to be $\sim 1$ femtobarn, possibly within reach of the high-luminosity LHC, with $\ge 140$ GeV of missing energy accompanied by two jets.}

\FullConference{%
  41st International Conference on High Energy physics - ICHEP2022\\
  6-13 July, 2022\\
  Bologna, Italy
}

\begin{document}
\maketitle

In earlier papers we proposed a new dark matter WIMP which has no interactions other than second-order interactions with W and Z bosons~\cite{DM2021a,DM2021b}. This particle is unique among viable dark matter candidates in that it has a well-defined mass and couplings, with no free parameters, so that in principle precise predictions can be made for all experimental cross-sections. As described below, the mass is determined by adjusting it to yield the observed dark matter relic abundance, with both this quantity and the cross-section for annihilation in the present universe calculated using MicrOMEGAs~\cite{MicrOMEGAs}. The cross-sections for direct detection and collider detection are estimated from results for the inert doublet model~\cite{Klasen-2013,Dutta}, by inferring the values in the limit that there is no Higgs coupling and that the masses of the particles other than the dark matter particle become extremely large: See Figs.~2 and 7 of Ref.~\cite{Klasen-2013}, and Figs.~2 and 10 (plus Table 3) of Ref.~\cite{Dutta}. (We have received communications from authors of these papers indicating that these estimates are reasonable. Now we are undertaking to replace the estimates by precise independent calculations.) 

All cross-sections are relatively low, and thus consistent with current experimental and observational limits, because they result only from gauge interactions which are second-order.

\begin{figure}
\begin{center}
\resizebox{0.45\columnwidth}{!}{\includegraphics{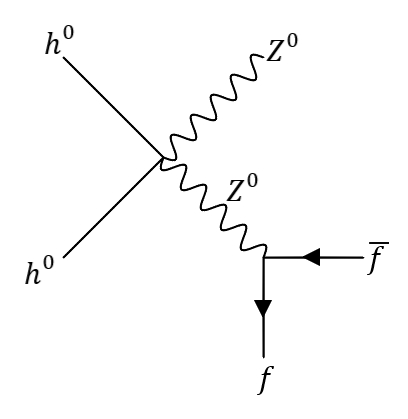}}
\resizebox{0.45\columnwidth}{!}{\includegraphics{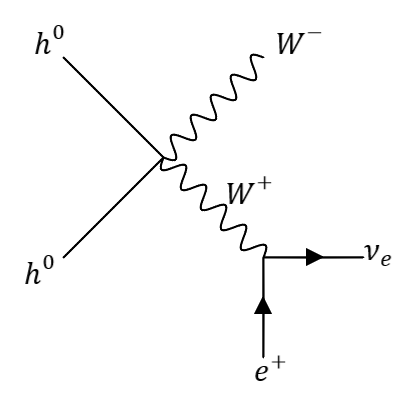}}
\end{center}
\caption{Representative diagrams for annihilation of the present dark matter candidate.}
\label{ann}
\end{figure}

In Fig.~\ref{ann} we show representative diagrams for annihilation of the present dark matter candidate -- the lowest-energy higgson\cite{DM2021a,DM2021b} -- if its mass $m_h$ is below the mass of the W boson. The currently preferred values of $\Omega_{DM} = 0.27$ and $h = 0.73$ imply that $\Omega_{DM} h^2= 0.144$. The calculations with MicrOMEGAs yield $\Omega_{DM} h^2= 0.134, 0.147, 0.162$ respectively for $m_h = 70.5, 70.0, 69.5$ GeV/c$^2$, so we conclude that $m_h \approx$  70 GeV/c$^2$ is required if the dark matter consists exclusively of this one component. 

It should be mentioned, however, that the present theory includes supersymmetry at some energy scale, and a lightest supersymmetric partner, such as the lightest neutralino, can stably coexist with the lightest higgson as a subdominant component, in a multicomponent scenario. Other components such as axions are also hypothetically possible, although such candidates have much more poorly defined masses and interactions.

\clearpage

For a mass of 70 GeV/c$^2$, our calculations yield a cross-section given by $\langle \sigma_{ann} v \rangle = 1.19 \times 10^{-26} $ cm$^3$/s for annihilation in the present universe. This value is consistent with the current limits from observations of dwarf spheroidal galaxies~\cite{Alvarez,Slatyer}. 

Both the mass and cross-section are also consistent with the interpretation that (i) the Galactic-center gamma ray excess observed by Fermi-LAT and (ii) the antiproton excess observed by AMS-02 result from annihilation of these particles in the present universe. The detailed analyses are cited as Refs~[34]-[43] in Ref.~\cite{DM2021a}, and a more extensive discussion and set of references is given in a recent Snowmass review~\cite{Snowmass2021}.
\begin{figure}
\begin{center}
\resizebox{0.45\columnwidth}{!}{\includegraphics{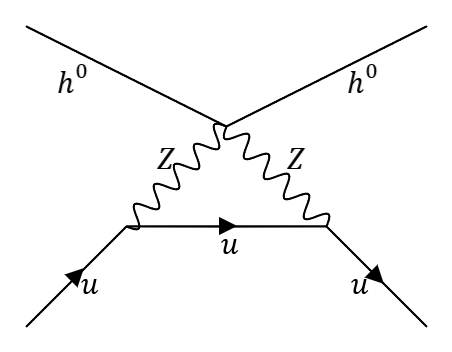}}
\resizebox{0.45\columnwidth}{!}{\includegraphics{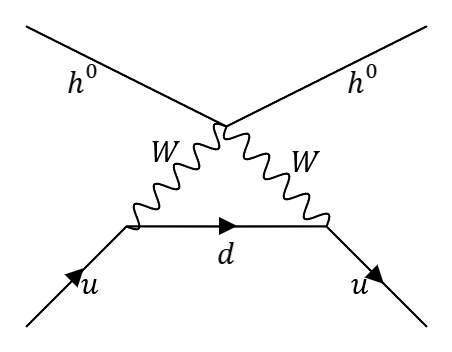}}
\end{center}
\caption{Representative diagrams for direct detection of the present dark matter candidate.}
\label{dir}
\end{figure}
\begin{figure}
\begin{center}
\resizebox{0.45\columnwidth}{!}{\includegraphics{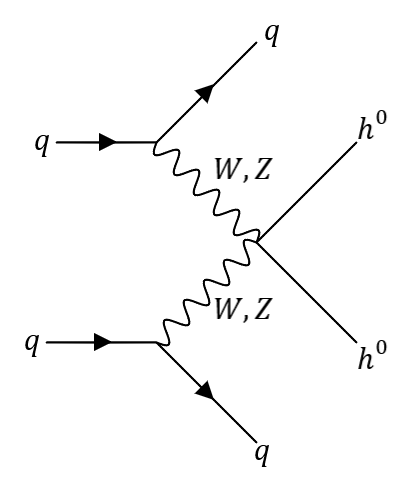}}
\resizebox{0.45\columnwidth}{!}{\includegraphics{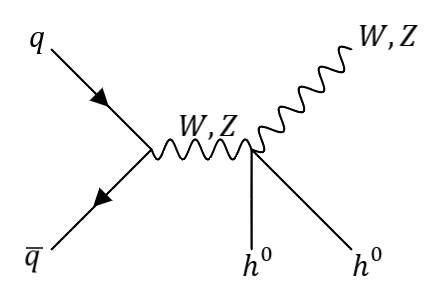}}
\end{center}
\caption{Representative diagrams for collider detection of the present dark matter candidate, with $\ge 140$ GeV of missing energy accompanied by two jets in the more promising vector-boson fusion process.}
\label{col}
\end{figure}

\clearpage

In Fig.~\ref{dir} we show representative diagrams for direct detection as the dark matter particle $h^0$ collides with the quarks in a nucleus. Based on the results of Ref.~\cite{Klasen-2013}, we estimate a cross-section which is slightly above $10^{-48}$ cm$^2$  in Xe-based experiments. This should be (barely) attainable within the next few years by LZ~\cite{LZ} and XENONnT~\cite{XENON}. In principle other direct-detection experiments, such as PandaX~\cite{PandaX} and (a repurposed) SuperCDMS, should be able to observe this particle on a longer time scale.

Fig.~\ref{col} shows representative diagrams for collider detection of the present dark matter candidate. Based on the results of~\cite{Dutta}, we estimate the cross-section for production via vector boson fusion to be $\sim 1$ femtobarn, which may be within reach of the high-luminosity LHC in 12-15 years, and other collider experiments on a longer time scale.

\end{document}